\documentclass[prb,preprint]{revtex4-1}

\usepackage{amsmath}  
\usepackage{amsfonts} 
\usepackage{graphicx} 
\usepackage{tabulary}

\begin{document}

\title{Interactive simulations for quantum key distribution}

\author{Antje Kohnle}
\email{ak81@st-andrews.ac.uk} 
\author{Aluna Rizzoli}
\affiliation{School of Physics and Astronomy, University of St Andrews, St Andrews, KY16 9SS, United Kingdom}

\date{\today}

\begin{abstract}
Secure communication protocols are becoming increasingly important, e.g. for internet-based communication. Quantum key distribution allows two parties, commonly called Alice and Bob, to generate a secret sequence of 0s and 1s called a key that is only known to themselves. Classically, Alice and Bob could never be certain that their communication was not compromised by a malicious eavesdropper. Quantum mechanics however makes secure communication possible. The fundamental principle of quantum mechanics that taking a measurement perturbs the system (unless the measurement is compatible with the quantum state) also applies to an eavesdropper. Using appropriate protocols to create the key, Alice and Bob can detect the presence of an eavesdropper by errors in their measurements.

As part of the QuVis Quantum Mechanics Visualization Project, we have developed a suite of four interactive simulations that demonstrate the basic principles of three different quantum key distribution protocols. The simulations use either polarized photons or spin 1/2 particles as physical realizations. The simulations and accompanying activities are freely available for use online or download, and run on a wide range of devices including tablets and PCs. Evaluation with students over three years was used to refine the simulations and activities. Preliminary studies show that the refined simulations and activities help students learn the basic principles of QKD at both the introductory and advanced undergraduate levels. 
\end{abstract}

\maketitle

\section{Introduction} 

Cryptography is the science of encrypting and decrypting messages for secure communication.\cite{Singh1999} It has been used for military purposes since ancient times. Cryptography is widely used today, e.g. for secure internet communication. The protocols used are based on classical physics and the assumption of limited computing power, which may one day turn out to be wrong!

The basic procedure of cryptography consists of a sender and a receiver, conventionally called Alice and Bob respectively, who wish to share a secret message despite the presence of a malicious eavesdropper, conventionally called Eve.\cite{Gisin2002} Alice and Bob can only communicate through public channels such as phone or internet. Thus, they need to encrypt their message prior to sending it as an eavesdropper could listen in on their communication. One simple way of encrypting the message is if Alice and Bob share a secret key. Such a key could be a random sequence of 0s and 1s, each of which is called a bit, only known to themselves. If the message is also in the form of a sequence of 0s and 1s, Alice can encrypt her message using bitwise addition of her message and the secret key, using addition modulo 2 where 1+1=0. For example, the message 11001 and the key 01001 give the encrypted message 10000. This encrypted message is called a ciphertext. Equipped with the same secret key, Bob now performs a bitwise addition of the ciphertext and the key to retrieve the original message. In the example, Bob performs a bitwise addition of the ciphertext 10000 and the key 01001 to retrieve the original message 11001. The procedure just described is called a one-time pad, and is only secure if the key is truly secret and used only once. Eve gains no information from intercepting the ciphertext, as for a ciphertext composed of $n$ bits there are $2^n$ equally likely messages that are compatible with the ciphertext.

How can Alice and Bob share the secret key needed for this procedure? This is the formidable challenge of key distribution. Classically, there is no way to ensure that the distribution of the key is absolutely secure. Quantum mechanics makes such secure key distribution possible! The fundamental principle of quantum mechanics that taking a measurement perturbs the system (unless the measurement is compatible with the quantum state) also applies to Eve. Thus, if Alice and Bob use appropriate protocols, Eve can not gain information about Alice's and Bob's communication without introducing perturbations that reveal her presence. Since Alice and Bob are using a quantum procedure to only create a key and not to send a message, they can simply disregard any key that has been compromised, keeping their communication secure.  Eve can not simply make a copy of the state that Alice sends to perform measurements on the copy, as the so-called no-cloning theorem proves that in the quantum world it is not possible to make a perfect copy of an arbitrary quantum state.\cite{Wootters1982}

The idea of quantum key distribution (sometimes imprecisely called quantum cryptography) was first developed by Bennett and Brassard in 1984 (see section \ref{sectionBB84}).\cite{BB84} Since then, the field has progressed enormously.\cite{LoReview2014} For example, quantum key distribution (QKD) using photons has been demonstrated over large distances using optical fibres\cite{Korzh2015} and in free space\cite{Wang2013}, and with systems that are stable over long periods of time\cite{Yoshina2013}. Commercial QKD devices are already available.  

Quantum mechanics courses often only discuss quantum information in advanced courses. We argue that aspects of quantum information can profitably be covered much earlier. Quantum key distribution (QKD) is one of the simplest demonstrations of the emerging and rapidly growing field of quantum information technology.  In the most common form (see section~\ref{sectionBB84}), it does not require complicated mathematics, and only requires a knowledge of basic quantum mechanics concepts such as superposition states and their measurement outcome probabilities, incompatible observables, the collapse of a quantum state on measurement, and a physical quantum system with two states as a realization of a qubit. It also links well with the spins-first or more generally two-level systems approach gaining favor in quantum mechanics instruction, where topics such as single photons or spin 1/2 particles that are physical realizations of qubits are covered early in the curriculum.\cite{Michelini2000, Bronner2009, Pearson2010, Scarani2010, Kohnle2014, Malgieri2014} While QKD relies on the no-cloning theorem, our experience is that QKD can be discussed at the introductory level without proving this theorem. QKD is a topic with direct everyday applications well known to students (such as secure internet communication), and thus helps to make a subject that can often appear abstract and far-removed from everyday experience relevant to students' lives. 

Despite distinct advantages of incorporating aspects of quantum information into quantum mechanics courses, many textbooks do not include this topic, and very limited numbers of multimedia resources and research-based materials exist compared with other quantum mechanics topics.\cite{Mason2014, SinghPERC2014} This article gives an overview of interactive simulations developed as part of the QuVis Quantum Mechanics Visualization Project\cite{Kohnle2014, Kohnle2015} to support the learning and teaching of the basic principles of quantum key distribution using three different protocols. The simulations use polarized single photons and spin 1/2 particles as physical realizations of qubits. Simulations can be used both at the introductory and advanced levels, and are also aimed at instructors with other areas of expertise interested in learning the principles of QKD. Simulations have been coded in HTML5/Javascript, so that they run on a wide range of devices including tablets and PCs. All of the QuVis simulations and accompanying activities are freely available for use online or download (www.st-andrews.ac.uk/physics/quvis). Instructors can download password-protected solutions to activities from the QuVis website. Instructors interested in obtaining the password for the solutions are requested to email quvismail@st-andrews.ac.uk. 

The simulations described here have a number of features that make them useful for learning about QKD. They allow students to experience how a raw key would be generated experimentally and to see the effect of an eavesdropper infiltrating the experiment. Students can easily set up different configurations, such as using only a single basis or two bases and see the results of data-taking immediately without needing complex equipment or long periods of time. They allow students to collect data at different speeds i.e.~via individual particles/photons or in fast-forward mode. They show simplified idealized systems, such as perfect sources and detectors and a simple intercept-resend eavesdropping attack, to focus on key ideas and reduce cognitive load. They help students make connections between physical representations of the experimental setup and mathematical representations, e.g.~the key bits, the error rate and the error probability. 

In what follows, we describe three different QKD protocols and show how they are implemented in the simulations. We then describe how the design of the simulations was refined using feedback from individual student volunteer sessions and in-class trials to ensure their educational effectiveness. We conclude with an outlook for future work.

\section{The BB84 protocol} \label{sectionBB84}

The best known quantum key distribution protocol is the so-called BB84 protocol, published by Bennett and Brassard in 1984.\cite{BB84}

In this protocol, Alice sends Bob a sequence of photons/particles in one of four different states chosen at random from two conjugate bases. For each photon/particle, Bob chooses one of the bases at random and makes a measurement in this basis. Once all measurements have been made, Alice and Bob exchange their bases (not their measurement outcomes!) and discard those measurements where their bases did not coincide. In order to test for an eavesdropper, they compare a randomly selected subset of their measurement outcomes (which they then discard, as these are not anymore secure) and check for errors. If the error rate is below a certain threshold, the unshared outcomes form the raw key, which can be processed further classically.

The \textit{Quantum key distribution (BB84 protocol) using polarized photons}\cite{BB84sim} simulation (see Fig.~1) uses single photons and photon polarization as a physical realization of the BB84 protocol. At startup, the simulation shows an Introduction view, with introductory text and a start button. The introductory text includes the learning goals, e.g. ``to help Alice and Bob decide whether or not they have generated a secure key. How can they tell that an eavesdropper Eve has infiltrated their experiment?''. This Introduction view is similar across all the QKD simulations. For the simulations using single photons, pressing the start button shows an animation depicting virtual reality goggles that allows students to ``see'' the photons in the simulation. The photon visualization was developed in an earlier study.\cite{KohnlePERC2014}

\begin{figure}[h!]
\centering
\includegraphics[width=16cm]{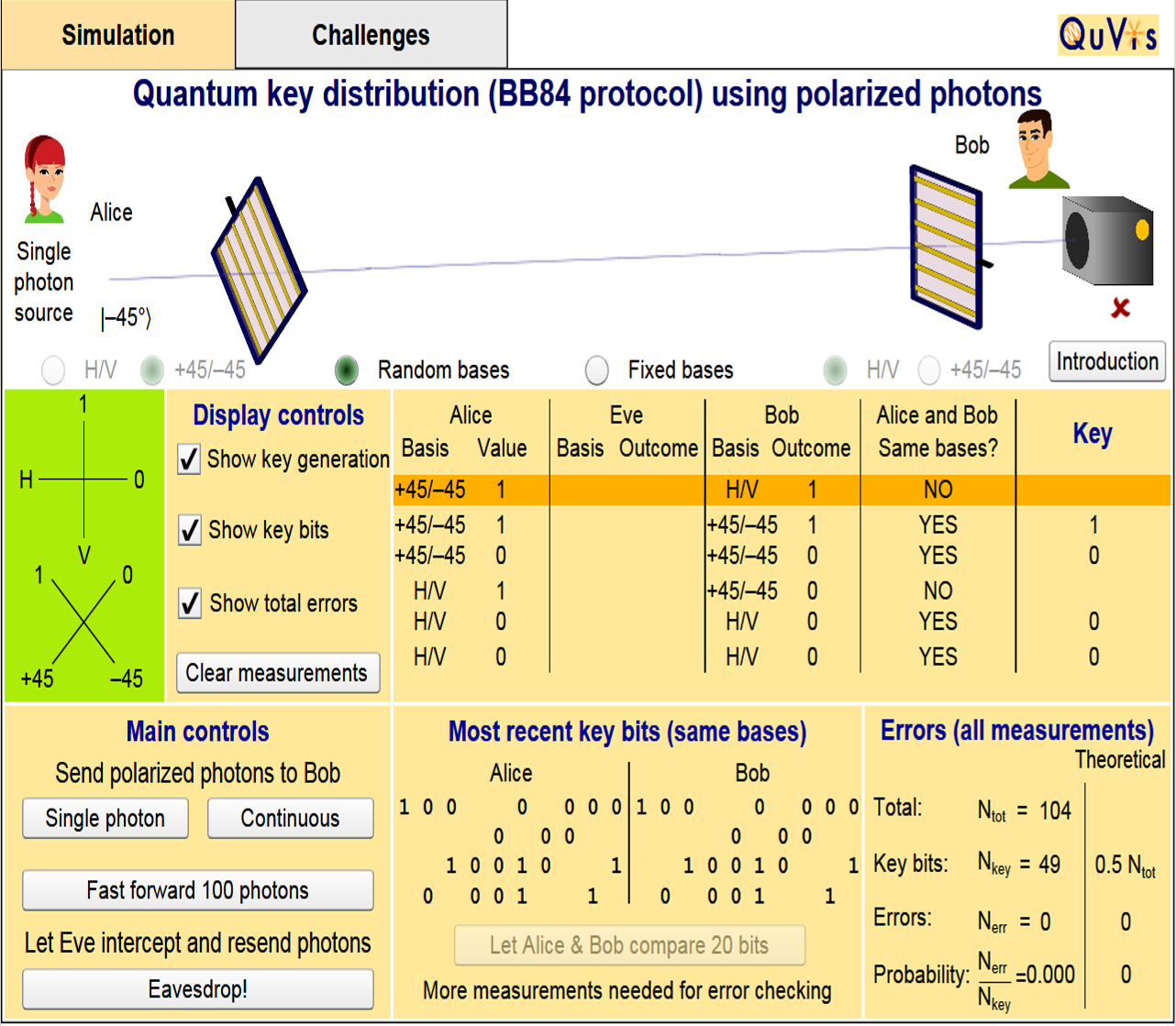}
\caption{A screenshot of the Controls view of the \textit{Quantum key distribution (BB84 protocol) using polarized photons} simulation with all the tickboxes checked, but no eavesdropper inserted. A test for errors has been previously carried out by Alice and Bob, as can be seen in the gaps in the middle bottom panel. These gaps correspond to shared bits that have been discarded as they are not anymore secure.}
\label{BB84screenshot}
\end{figure}

A screenshot of the Controls view of the BB84 simulation is shown in Fig.~1. As shown in the top graphics panel, Alice is equipped with a single photon source and a polarizer. She sends Bob a sequence of photons randomly polarized along either the horizontal, vertical, $+45$ or $-45$ directions, i.e. each in one of the quantum states $|H\rangle$, $|V\rangle$, $|+45\rangle$ or $|-45\rangle$. The horizontal and vertical directions constitute the $H/V$ basis, the $+45$ and $-45$ directions the $+45/-45$ basis. The two bases are chosen such that the two states in one basis are an equal superposition of the two states in the other basis, e.g. $|+45\rangle = 1/\sqrt{2} \left(|H\rangle + |V\rangle \right)$ and $|-45\rangle = 1/\sqrt{2} \left(|H\rangle - |V\rangle \right)$.
 
Bob is equipped with a polarization analyzer and a single photon detector. For each measurement, Bob chooses one of the two bases at random and and measures the polarization of the photon. In the simulation, it is assumed that Alice tells Bob each time she sends a photon. Bob either registers a photon in his detector or not, determining his outcome. For example, if Bob orients his analyzer horizontally but does not detect a photon, his outcome is ``vertical''; if he does detect a photon, his outcome is ``horizontal". In real experiments, polarizing beam splitters and two detectors would be used, so that each orientation would have its own detector and Bob would detect the photon in both cases. 
 
If by chance Bob chooses the same basis as Alice and no eavesdropper is present, his measurement outcome will be identical to what Alice sent, as the simulation assumes perfect sources and detectors. If by chance Bob chooses a different basis to Alice, either of his outcomes are equally likely, and his measurement has no information on the state that Alice sent. For example, if Alice sends $|V\rangle$ and Bob measures in the $H/V$ basis, he will measure $|V\rangle$, but if he measures in the $+45/-45$ basis, both outcomes $|+45\rangle$ and $|-45\rangle$ are equally likely.  Thus, whenever Alice and Bob use the same basis, they know that their measurements agree without needing to communicate these outcomes! The measurement outcomes are turned into bits by assigning the vertical and $-45$ orientations the value 1, and the horizontal and $+45$ orientations the value 0. 

As can be seen in Fig.~1, the simulation shows the basis and outcome for each individual measurement for the last five measurements with the current measurement highlighted (middle right panel), the most recent key bits (middle bottom panel) and the number of key bits and errors for all measurements taken for a given configuration (right bottom panel). The most recent key bits panel only includes those measurements for which Alice and Bob used the same basis, as only these measurements are used for the key. 

In the simulation, Eve uses a simple intercept-resend attack where she intercepts each photon sent to Bob, measures its polarization, and passes on a photon to Bob. If random bases are used, Eve orients her polarizer at random, and then passes on a photon with the polarization state she measured to Bob. If Eve by chance uses the same basis as Alice, she passes on Alice's state unchanged. If however, she used the incorrect basis, her measurement changes the quantum state, so that Bob has a 50\% chance of measuring either outcome. This latter case can lead to errors, e.g. if Alice sends $|V\rangle$ and Bob measures $|H\rangle$ and thus their bit values disagree even though they used the same basis. As Eve guesses the incorrect basis 50\% of the time, and 50\% of these cases lead to an error, the presence of Eve leads to 25\% of the key bits being incorrect on average. 

In order to test for an eavesdropper, Alice and Bob compare a randomly selected subset of their measurement outcomes. They can detect the presence of an eavesdropper by errors in their measurements. If they do not find errors, they discard the exchanged bits, with the remaining bits forming the raw key. Via the most recent key bits panel (middle bottom panel in Fig.~1), students can let Alice and Bob carry out a test for errors by comparing 20 bits.  In the view in Fig.~1, such a test has been carried out and the compared bits have then been removed (seen as the gaps in the middle bottom panel) as they are not anymore secure. 

The Controls view of all the QKD simulations is designed with implicit scaffolding to help students progress from simpler to more complex situations. At startup, the ``Fixed bases'' button in the top graphics panel is selected, so that only a single basis is used, and no eavesdropper is inserted. The tickboxes are all unticked, so that only the top graphics window, the display controls and the main controls are visible. This leads to students typically first exploring the simpler situation of a single basis before progressing to two bases. Students can see that when Eve is eavesdropping and only a single basis is used, Eve knows the entire key despite there being no errors in Alice's and Bob's measurements. If students let Alice and Bob compare 20 bits for errors, the simulation gives feedback ``No errors, but fixed bases, so not secure''. Thus, the one-basis case helps students recognize why two bases rather than a single basis are needed for secure QKD.  

The QuVis collection includes a second simulation demonstrating the BB84 protocol. The \textit{Quantum key distribution (BB84 protocol) with spin 1/2 particles}\cite{BB84spin} simulation uses spin 1/2 particles and the component of spin as determined with a Stern-Gerlach apparatus as a physical realization of the BB84 protocol. The four polarization states above correspond to the spin states spin-up and spin-down along the vertical (denoted as $|\uparrow\rangle$ and $\downarrow\rangle$), and spin-up and spin-down along the horizontal (denoted as $|+\rangle$ and $-\rangle$). 

\section{The two-state protocol} \label{sectionB92}
In 1992 Bennett demonstrated that four states are more than are actually needed for QKD.\cite{Bennett92} QKD requires at least two states, so that Eve is unable to unambiguously distinguish between them without introducing some errors in the key. However, Bennett demonstrated that two states are also sufficient if they are not orthogonal e.g. using photon polarization states $|H\rangle$ and $|+45\rangle$.

The \textit{Quantum key distribution using two non-orthogonal states}\cite{B92sim} simulation (Fig.~2) demonstrates this two-state protocol using single photons and photon polarization for the physical realization. Similarly to the BB84 simulation, Alice sends polarized photons to Bob, who is equipped with a polarization analyzer and single photon detector. However, Alice now uses two non-orthogonal polarization states rather than four polarization states. In the simulation, Alice randomly prepares each photon with either $0^\circ$ (horizontal) or $+45^\circ$ polarization. The horizontal polarization is assigned a bit value of 0, the $+45$ polarization a bit value of 1.  

\begin{figure}[h!]
\centering
\includegraphics[width=16cm]{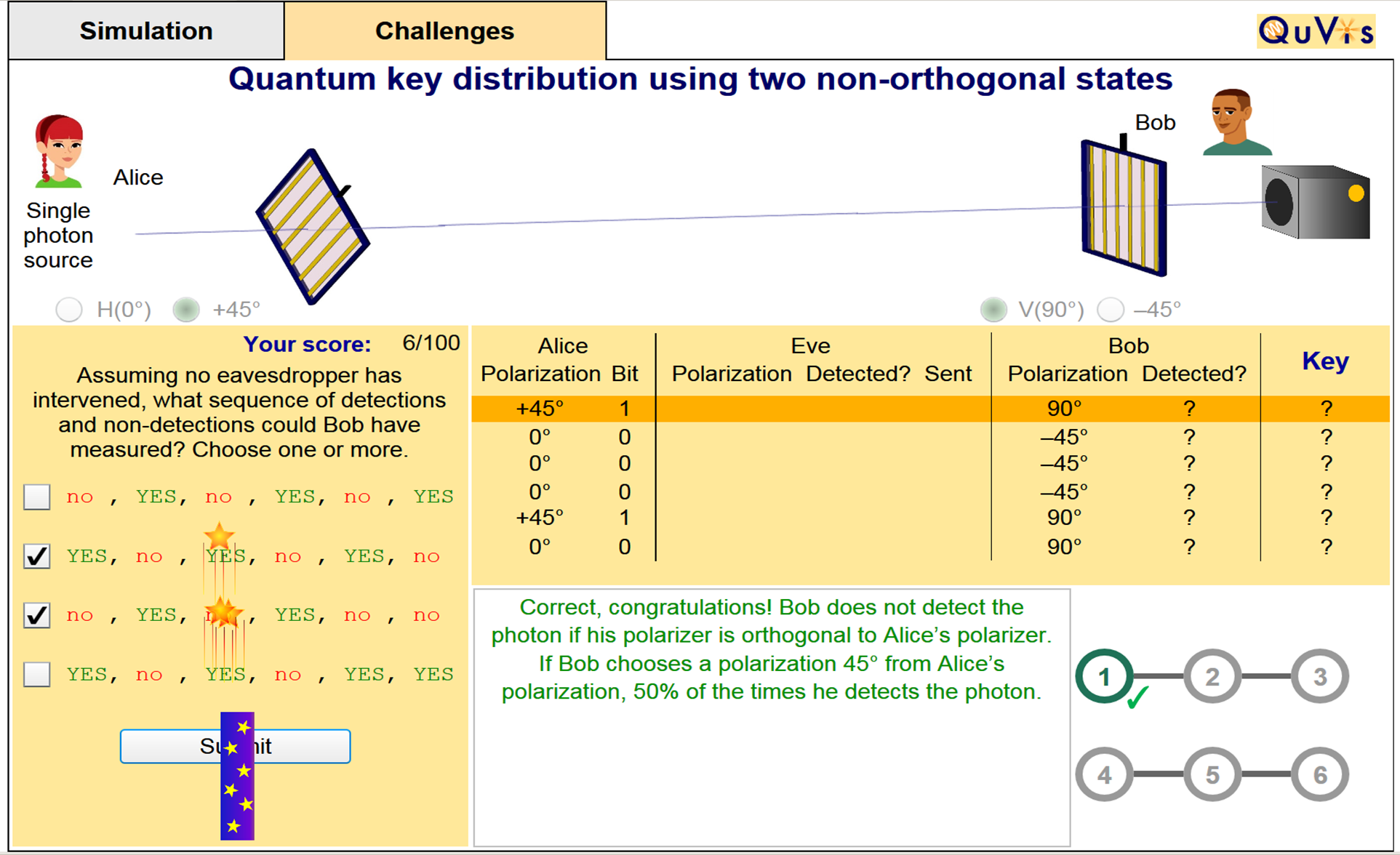}
\caption{A screenshot of the Challenges view of the \textit{Quantum key distribution using two non-orthogonal states} simulation. Challenges can be solved in any order, include a score counter, and provide feedback on correct and incorrect choices.}
\label{B92screenshot}
\end{figure}

For each measurement, Bob randomly sets his analyzer to one of two directions orthogonal to Alice's directions, so either $90^\circ$(vertical) or $-45^\circ$. Alice informs Bob whenever she sends a photon. If Bob detects the photon, he knows with certainty the polarization and hence the bit value (0 or 1) sent by Alice. For example, if Bob detects a photon when measuring along $90^\circ$, he knows that Alice sent a photon with $+45^\circ$ polarization (it cannot have been the $0^\circ$ polarization) and thus with bit value 1. Bob can therefore assign detections with $90^\circ$ a bit value of 1, and detections with $-45^\circ$ a bit value of 0. 

If Bob does not detect the photon, he cannot be certain which state Alice sent. For example, if Bob measures along $90^\circ$ and did not detect a photon, Alice could have sent either $0^\circ$ or $+45^\circ$. Thus, Alice and Bob keep only those measurements where Bob detected a photon. For example, if Alice sends $0^\circ$, then in 50\% of cases Bob measures along $90^\circ$ and does not detect the photon, and in 50\% of cases he measures along $-45^\circ$, with a 50\% chance of detection. Thus, key bits are produced in the simulation only in 25\% of cases. Alice and Bob publicly communicate to determine which photons were detected. This sequence of 0s and 1s forms the raw key. As for the other QKD protocols, they then exchange a small number of their bit values (which they then discard as they are not anymore secure) to check for errors.

In the simulation, Eve uses an intercept-resend attack where she uses the same orientations as Bob for her polarization measurement, and the same orientations as Alice for her photon source. If Eve detects the photon, she knows what state Alice sent and sends the same state as Alice on to Bob. In this case, errors do not occur. If Eve does not detect a photon, she does not know which state Alice sent. Eve then sends the state that is orthogonal to her polarizer as this is the more likely case. If Eve chooses the incorrect state to pass on to Bob, errors can occur. For example, if Alice sends $|H\rangle$ and if Eve measures along $-45^\circ$ and does not detect a photon and hence sends $+45^\circ$, then Bob can detect the photon in state $|V\rangle$, leading to an error.  The error probability is 25\%, the same as for the BB84 protocol. 

Excepting the \textit{Quantum key distribution with entangled spin 1/2 particles} simulation discussed in section \ref{sectionBBM92}, all the QKD simulations include a second Challenges tab with multiple challenges aligned with the learning goals. Challenges can be solved in any order, and include a score counter and green ticks showing which challenges have already been solved. Fig.~2 shows one of the challenges for the \textit{Quantum key distribution using two non-orthogonal states} simulation. This challenge asks students what sequence of detections and non-detections would be compatible with the given polarizations of Alice and Bob. Other challenges ask students what sequence of key bits could have been measured given Bob's outcomes, whether Eve detected a photon and what state she sent on to Bob for a particular measurement where Alice and Bob find an error, and whether Eve's analyzer orientation can be known for a case where Alice and Bob do not find an error.

\section{The Einstein-Podolsky-Rosen protocol} \label{sectionBBM92}

In 1991, Ekert suggested a QKD protocol using entangled particle pairs rather than individual particles and measurements along three axes.\cite{Ekert1991} This scheme was simplified by Bennett, Brassard and Mermin in 1992 to measurements along two orthogonal axes.\cite{BBM92} This simplified scheme is implemented in the simulation described here.

\begin{figure}[h!]
\centering
\includegraphics[width=16cm]{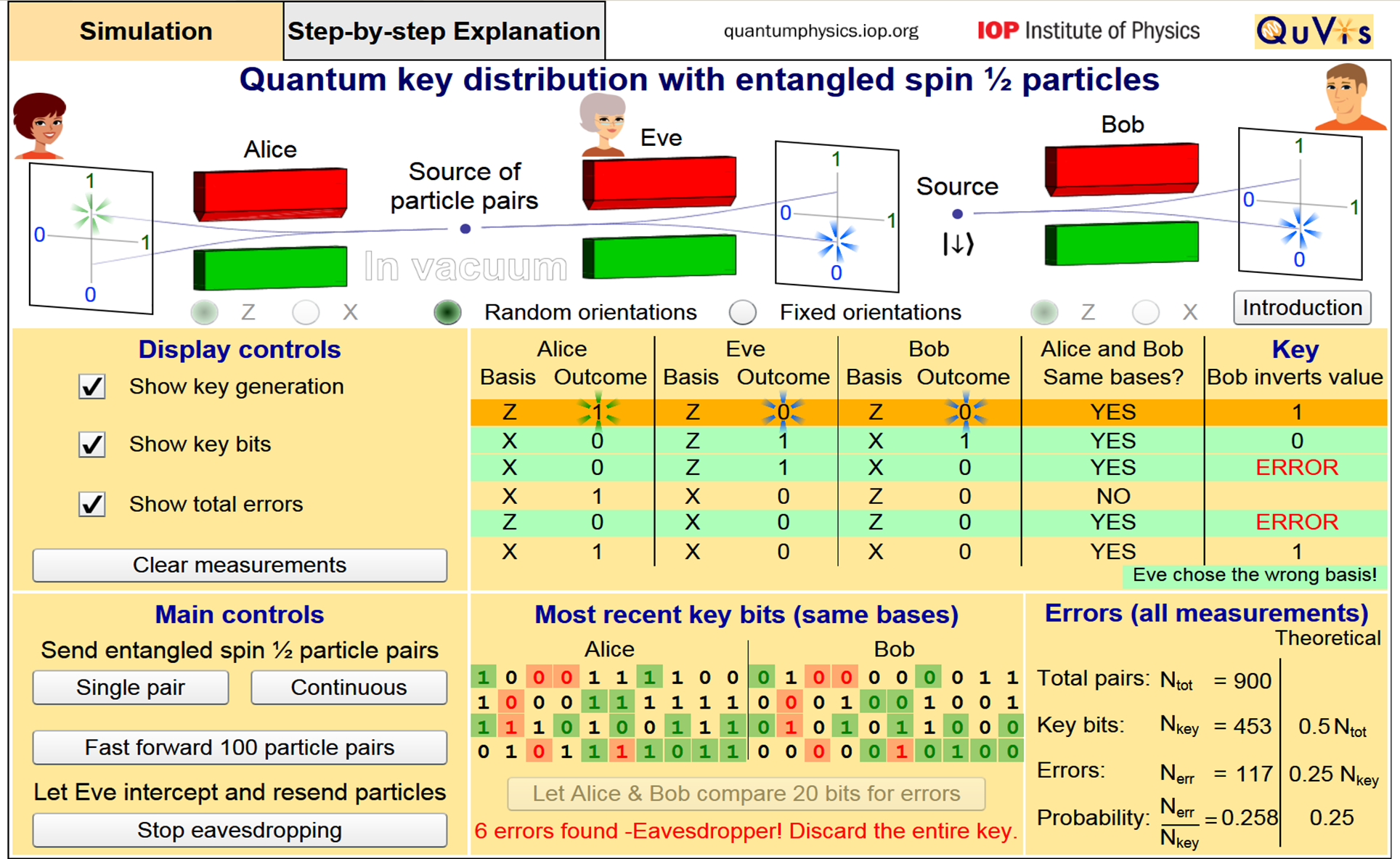}
\caption{A screenshot of the Controls view of the \textit{Quantum key distribution with entangled spin 1/2 particles} simulation with all the tickboxes checked and the eavesdropper inserted. A test for errors has been carried out in the middle bottom panel, with incorrect values highlighted in red.}
\label{BBM92screenshot}
\end{figure}

The \textit{Quantum key distribution with entangled spin 1/2 particles}\cite{BBM92sim} simulation uses a source that emits entangled spin 1/2 particle pairs (see Fig.~3).  In the simulation, the two particles in the pair are emitted back-to-back each with opposite spin components. The particle pair is in the maximally entangled quantum state
$|\psi_{AB}\rangle = 1/\sqrt{2} \left( |\uparrow_A\rangle |\downarrow_B\rangle - |\downarrow_A\rangle |\uparrow_B\rangle \right)$, where $|\uparrow\rangle$ and $|\downarrow\rangle$ correspond to spin-up and spin-down states along the vertical and the indices A and B stand for Alice and Bob. 

As can be seen in the top panel of Fig.~3, Alice and Bob are each equipped with a Stern-Gerlach apparatus, which uses two magnets (one of which is pointed) to create a region of non-uniform magnetic field aligned along a given axis.  For spin 1/2 particles, the particles separate into two discrete streams, one deflected in the positive direction (in the simulation defined as measurement outcome 1), one deflected in the negative direction (measurement outcome 0) along this axis.  Alice and Bob orient their Stern-Gerlach apparatuses independently of one another at random along two orthogonal axes, denoted X and Z.

For each particle pair, Alice and Bob note the basis (X or Z) and measurement outcome (0 or 1). Due to the maximally entangled state used, they know that their measurement outcomes are perfectly anticorrelated (if Alice measures 1, Bob measures 0 and vice versa) for those measurements where both Stern-Gerlach apparatuses happened to be oriented along the same axis. In contrast to a classical anticorrelation, this perfect anticorrelation of the entangled state holds both along the X and the Z bases. This can be seen mathematically using the transformations $|\uparrow\rangle = 1/\sqrt{2} \left(|+\rangle +|-\rangle \right)$ and
$|\downarrow\rangle = 1/\sqrt{2} \left(|+\rangle - |-\rangle \right)$, where $|+\rangle$ and $|-\rangle$ correspond to spin-up and spin-down states along the horizontal respectively. If Alice and Bob use different bases (e.g. X and Z), then their outcomes are completely uncorrelated. Thus, as for the BB84 protocol, Alice and Bob publicly share the bases used for each measurement but not their measurement outcomes, and keep only those outcomes for which their bases were the same.  

Alice and Bob then exchange a small number of their actual measurement outcomes (which they then discard from the key as they are not anymore secure) to check for errors. As for all the QKD simulations, students can insert an eavesdropper Eve who intercepts the particles sent to Bob (shown in Fig.~3). For the "Random orientations" setting, Eve measures the particle's spin component in the same way as Bob using a Stern-Gerlach apparatus oriented at random along X or Z. She then sends a particle on to Bob with the spin state she measured, e.g. if Eve measures spin-up along Z, she sends a particle in the state $|\uparrow \rangle$ to Bob.  

Errors occur in 50\% of cases where Eve by chance chooses the wrong basis, i.e. a different one to Alice and Bob. As Alice and Eve have used different bases, their outcomes are completely uncorrelated. Eve's measurement changes the spin state of the particle. Eve then passes on a particle to Bob that has equal probabilities for outcomes 0 and 1 in his basis. As Eve guesses the incorrect basis 50\% of the time, and 50\% of these cases lead to an error, the quantum bit error rate is 25\%, the same as for the BB84 and two-state protocols.

Fig.~3 shows the setup with random orientations and Eve inserted. In the middle right key panel, one can see measurements highlighted where Eve has chosen the wrong basis. One can see that errors occur only sometimes, not always, if Eve used a different basis to Alice and Bob (compare lines 2, 3 and 5 of this panel). In the middle bottom panel, a test for errors has just been carried out. The 20 bits that were compared are highlighted, with errors highlighted in red. The errors indicate the presence of an eavesdropper, so that the entire key needs to be discarded.

\section{Assessing simulation effectiveness} \label{effectiveness}
We developed an initial version of the \textit{Quantum key distribution with entangled spin 1/2 particle pairs}\cite{oldBBM92sim} simulation (coded in Adobe Flash) in 2013. After development, we ran a small number of individual student volunteer sessions where students interacted with the simulation freely while thinking aloud and then worked on the associated activity. This led to minor revisions of the simulation such as the inclusion of a fast forward button, adding images of Alice, Bob and Eve, and minor revisions to texts and layouts. After incorporation of these revisions, this version was used in 2014 at the University of St Andrews in an introductory quantum physics course (similar to a US Modern Physics course) and in an advanced quantum mechanics course taken by students in their third or fourth year of study. Table~\ref{datatable} summarizes data collection from 2014 and 2016 (the latter described below). The 2014 version of the simulation had two differences to the current version described in section \ref{sectionBBM92} and used in the 2016 trial: instead of the middle bottom panel showing recent key bits and allowing a test to compare for errors (see Fig.~3), this simulation showed a graph of the error probability versus the number of key bits. This graph showed how the error probability converges to 0.25 in the limit of a large number of measurements, if random orientations are chosen and the eavesdropper is inserted. The other difference was in the key bits panel, which did not highlight measurements where Eve chose the wrong basis (the turquoise highlighting of the middle right panel in Fig.~3). In what follows, we motivate why these two changes were incorporated into all the QKD simulations described in sections \ref{sectionBB84}, \ref{sectionB92} and \ref{sectionBBM92} in light of the 2014 results.

\begin{table}[h!]
\centering
\caption{Evaluation carried out with two versions of the \textit{Quantum key distribution with entangled spin 1/2 particle pairs} simulation.}
\begin{ruledtabular}
\begin{tabular}{l c c c c }
Year & Level & N & Simulation/Activity version & How used \\
\hline	
2014 & introductory & 73 (activity)  & initial version & homework assignment\\
 &  & 60 (post-test) & & post-test in class \\
2014 & advanced & 14 (activity  & initial version & homework assignment \\
 &  & and post-test) &  &  post-test in class \\
2016 & introductory & 82 (activity  & revised version & PC classroom \\
 &  & and post-test) &  & workshop \\
\end{tabular}
\end{ruledtabular}
\label{datatable}
\end{table}

In both the 2014 courses the simulation activity was given as a homework assignment. Students had one week to complete the assignment which did not contribute to the course grade. In the lecture following the submission date, students were asked to complete a short post-test, which was identical for both levels. Students in the introductory course were only given the definition of a key in the lecture and otherwise did not discuss ideas of cryptography during class time, so that students had essentially no prior knowledge and were learning about QKD from the simulation. At the advanced level, this particular protocol was not discussed during class time, but the BB84 protocol had been discussed in detail.

We coded each of the activity responses as correct, partially correct, incorrect and unanswered. All percentages quoted below refer to fully correct answers. On the whole, the 2014 activity questions were well answered (N=73, 73.3\% correct for the introductory level, and N=14, 92.0\% correct for the advanced level), with however two exceptions seen at both levels. 

The first common difficulty seen was in response to the question ``On average, what fraction of Bob's key will be incorrect when Eve intercepts and resends particles? Explain how this fraction come about.''. The explanations for how errors come about were not always correct, with 41.1\% correct at the introductory level and 71.4\% at the advanced level. The most common incorrect idea seen at both levels was that errors occur every time Eve chooses the wrong basis. This is incorrect, as only 50\% of these cases lead to errors. For example, an introductory level student stated \textit{``About 1/4 of Bob's key will be incorrect. These errors occur when Bob's SGA [Stern-Gerlach apparatus] has the same alignment as Alice's (1/2 of the time), and Eve has a different alignment than Alice's (1/2 the time) so the chances of both happening together are 1/4.''} and an advanced level student similarly stated \textit{``On average, 1/4 of Bob's key will be incorrect. 1/2 probability due to his choice of basis. 1/2 due to Eve's choice of basis. 1/2 $\times$ 1/2 = 1/4.''}  These responses are incorrect in that they do not consider that errors refer only to key bits, i.e., those measurements where Alice and Bob chose the same basis. Thus, the factor of 1/2 for Alice and Bob choosing the same basis is not relevant for the error probability. These responses are also incorrect in only considering the bases, not the measurement outcomes. If Eve chooses the wrong basis, e.g. X rather than Z, then her measurement projects the state into an eigenstate of X. Thus, Bob has equal probabilities of measuring a 0 or a 1 along Z. Eve chooses the wrong basis 50\% of the time. In these cases, Bob then has a 50\% chance of measuring an error, leading to the 25\% error fraction. 

The second common difficulty seen was in response to the question ``What actions could Alice and Bob take to determine whether or not Eve has compromised the security of their key?''. This question was not well answered, with 58.9\% correct at the introductory level (N=73) and 78.6\% at the advanced level (N=14). All answers in terms of Alice and Bob checking their key bits for errors were marked as correct.  At the introductory level, 9.6\% of students stated that Alice and Bob could exchange a message and see whether it is transmitted satisfactorily. 28.8\% of students made other incorrect suggestions not related to the protocol, such as Alice and Bob taking measurements with their Stern-Gerlach apparatuses set to the same orientation. The other incorrect answers made incorrect reference to the protocol. The advanced level students were presumably answering this question correctly more often based on their prior knowledge of the BB84 protocol.  

These two common difficulties pointed to shortcomings of the 2014 version of the simulation: it did not sufficiently show that errors do not occur every time Eve chooses a different basis to Alice and Bob, and did not demonstrate how Alice and Bob could test for errors. Therefore, we replaced the error probability graph by the most recent key bits panel shown at the bottom middle of Fig.~3, that lets Alice and Bob compare 20 bits for errors. Students can see that they need to discard any bits that were shared (the gaps in this panel in Fig.~1), and that if an eavesdropper is detected they need to discard the entire key (the feedback in this panel in Fig.~3). This panel is available in all the QKD simulations described in sections \ref{sectionBB84}, \ref{sectionB92} and \ref{sectionBBM92}. We also introduced the turquoise highlighting shown in the middle right panel in Fig.~3, that highlights all measurements where Eve chose the wrong basis. This allows students to see that not all of these measurements lead to an error (e.g. of the three measurements highlighted in this way in Fig.~3, only two of them have an error).  We also modified the activity, to extend the focus on how errors come about using two additional questions that asked students to explain using a specific example how an error occurs, what they can say about Eve's orientation when an error occurs and whether an error occurs each time Eve chooses the wrong orientation. 

We used the revised version of the \textit{Quantum key distribution with entangled spin 1/2 particle pairs} simulation shown in Fig.~3 and the revised activity in 2016 in the same introductory quantum physics course. Students completed the simulation assignment in a computer classroom workshop. In the 50-minute long workshop, students interacted with the simulation and worked on the assignment for typically 35-40 minutes with facilitator support. In the last minutes of the workshop, students completed the same post-test as in 2014. Students were asked to close the simulation and work alone when completing the post-test. 

Table~\ref{outcometable} gives a comparison of results for the two activity questions across both years that relate to the difficulties described above. Both of these questions were answered significantly better in 2016 compared with 2014 (U=1887, N1=82, N2=73, p$<$0.005 for the error fraction question, U=2420, N1=82, N2=73, p=0.01 for the actions to detect the presence of Eve question). In 2016, 78.0\% of students answered both questions correctly, compared with 41.1\% and 58.9\% respectively in 2014. This indicates that the revisions incorporated into the simulation and activity were effective in helping students come to a better understanding of how the error fraction of 25\% comes about and how Alice and Bob go about checking for errors. A typical correct response from the 2016 data describing how the error fraction comes about was \textit{``0.25 [error probability] since if Eve is orientated the same as Bob, no error will occur, this happens 50\% of the time. If Eve is orientated differently, then there is a 50\% chance an error will occur, so 50\% of 50\% is 25\% so a 1/4 chance Bob's key will be incorrect.''}

\begin{table}[h!]
\centering
\caption{The percentage of correct answers for two identical activity questions used in the 2014 and 2016 in-class trials. These two questions relate to changes in the simulation (for the error fraction also to the activity) implemented between the 2014 and 2016 trials.}
\begin{ruledtabular}
\begin{tabular}{l c c }
Activity question  &  Percent correct 2014 &  Percent correct 2016  \\
(abbreviated) & introductory level (N=73) & introductory level (N=82) \\
\hline	
Explain how the error fraction  & 41.1\% & 78.0\% \\
comes about.  &  & \\
What actions can Alice and Bob  & 58.9\% & 78.0\% \\
take to determine whether or not  &  & \\
Eve has compromised their key?  &  & \\
\end{tabular}
\end{ruledtabular}
\label{outcometable}
\end{table}

The post-test questions are given in the Appendix. Results for both years for the introductory level are shown in Fig.~4. The results show that the majority of students are achieving the learning outcomes in terms of understanding which measurements are used for the key and how errors come about in the presence of an eavesdropper. The results for questions 1 and 3 are the same within errors across both years. Question 2 is answered significantly better in 2016 than in 2014 (U=2127.5, N1=82, N2=60, p=0.02). The post-test results for the 2014 advanced level are not shown in Fig.~4. They were 100\%, 92.9\% and 85.7\% respectively for the three questions (N=14), but students likely had substantial prior knowledge so that the outcomes do not relate to learning from the simulation alone. 

In summary, these evaluation results point to the simulation being effective in helping students make sense of QKD at both the introductory and advanced levels. They point to the revisions incorporated into the QKD simulation and activity being successful in improving students' understanding. 

\begin{figure}[h!]
\centering
\includegraphics[width=12cm]{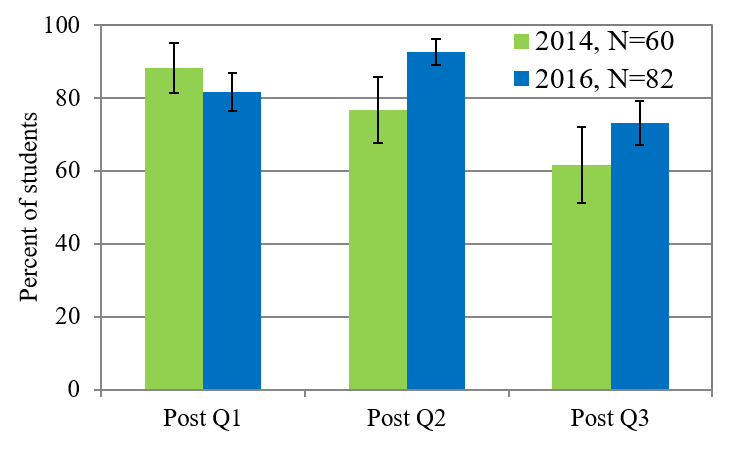}
\caption{Post-test outcomes for the 2014 and 2016 in-class trials of the \textit{Quantum key distribution with entangled spin1/2 particle pairs} simulation at the introductory level. Error bars represent the standard error of a proportion. The post-test questions are given in the Appendix. }
\label{posttest-figure}
\end{figure}

\section{conclusions}
QKD is a relevant and useful topic for inclusion in quantum mechanics courses at both the introductory and advanced levels, in terms of limited prior quantum mechanics knowledge needed, the rapidly growing field of quantum information technology and the link to motivating real-world applications. We have described four interactive simulations that support the learning and teaching of QKD. The simulations demonstrate the basic principles of QKD using three different protocols, and using either polarized single photons or spin 1/2 particles as physical realizations of qubits. The simulations have been shown in preliminary studies to be effective at helping students learn about QKD both at the introductory and the advanced undergraduate levels. 

The simulations described aim to only demonstrate the basic principles of QKD, and thus show idealized and simplified situations, e.g. assuming perfect sources and detectors so that no errors occur without the presence of an eavesdropper. The simulations also assume that Eve uses a simple intercept-resend strategy and listens in on each bit.
Future work will aim to develop simulations that include imperfect detectors and other possible eavesdropping strategies, such as Eve only intercepting a fraction of particles. Due to instrumental imperfections, some errors would always be present. If the error rate is not too large, Alice and Bob can use classical post-processing to create a secure key.\cite{VanAssche2006} This consists of error-correction methods to reduce errors between their two raw keys and privacy amplification to reduce Eve's information on Alice's bits, both at the expense of reducing the length of the key. Future work will aim to include in the simulations errors due to instrumental imperfections and classical post-processing. This will allow the simulations to demonstrate how a secure key can be generated even in the presence of some errors.

\appendix*   

\section{Post-test questions}

Consider secure key generation with entangled spin 1/2  particle pairs as shown in the simulation. The two particles in the pair are described by a single wavefunction 
$|\psi_{AB}\rangle = 1/\sqrt{2} \left( |\uparrow_A\rangle |\downarrow_B\rangle - |\downarrow_A\rangle |\uparrow_B\rangle \right)$. Assume observers Alice and Bob each have a Stern-Gerlach apparatus which they can orient independently and randomly along two orthogonal axes, X and Z. 

Alice and Bob make measurements on 5 pairs in order to generate a secure key. Bob uses the orientations X X Z X Z (in this order) for his measurements and obtains the results 01100 (a deflection in the positive direction is 1, and in the negative direction 0). Alice tells him he measured the first, second and last in the ``correct'' orientation (so in the same orientation as Alice).

\noindent 1) How many bits are there in Alice's and Bob's shared key? \\
\noindent A) 1 \hspace{1cm}	B) 2	\hspace{1cm} C) 3	\hspace{1cm}  D) 4	\hspace{1cm}	E) 5 

\noindent Alice and Bob decide to compare all of their bits to determine if Eve was intercepting. They find that they do not agree on the last bit of the key.

\noindent 2) For this last bit, what orientation must Eve have used for her measurement? \\
\noindent A) X	\hspace{1cm} 	B) Z \hspace{1cm} 	C) It is not possible to tell from the information given.

\noindent 3) For this last bit, what value did Eve obtain in her measurement? \\
A) 0		\hspace{1cm}  1) 1		\hspace{1cm}   C) It is not possible to tell from the information given.

\begin{acknowledgments}

We thank the UK Institute of Physics and the University of St Andrews for funding the simulation development. We thank Martynas Prokopas for coding the 2014 version of the quantum key distribution simulation. We thank Holly Scott-Riddell and Mark Paetkau for work on the data analysis. We thank Seth DeVore for suggestions for improvement of the \textit{Quantum key distribution using two non-orthogonal states} simulation. We thank Charles Baily and Natalia Korolkova for implementing the quantum key distribution simulation in their courses. We are grateful to all students taking part in evaluation studies and providing feedback on simulations. 

\end{acknowledgments}

\end{document}